\documentclass{ws-p10x7}
\newcommand{\lsim}{\raisebox{0.5mm}{\em $\, <$} 
\hspace{-2.8mm} \raisebox{-1.3mm}{\em $\sim \,$}}
\begin{document}

\title{Four neutrino oscillation analysis of atmospheric neutrino data
and application to long baseline experiments}

\author{Osamu Yasuda}

\address{Department of Physics, Tokyo Metropolitan University\\
        1-1 Minami-Osawa Hachioji, Tokyo 192-0397, Japan\\
E-mail: yasuda@phys.metro-u.ac.jp}

\twocolumn[\maketitle\abstract{
Analyis of the Superkamiokande atmospheric neutrino data
is presented in the framework of four neutrinos
without imposing constraints of Big Bang Nucleosynthesis.
Implications to long baseline experiments are briefly discussed.
}]

\section{Four neutrino analysis of atmospheric neutrinos}
\begin{figure}[h]
\vglue -1.35cm
\hglue -2.5cm
\hbox to\hsize{\hss\epsfxsize=6.2cm\epsfbox{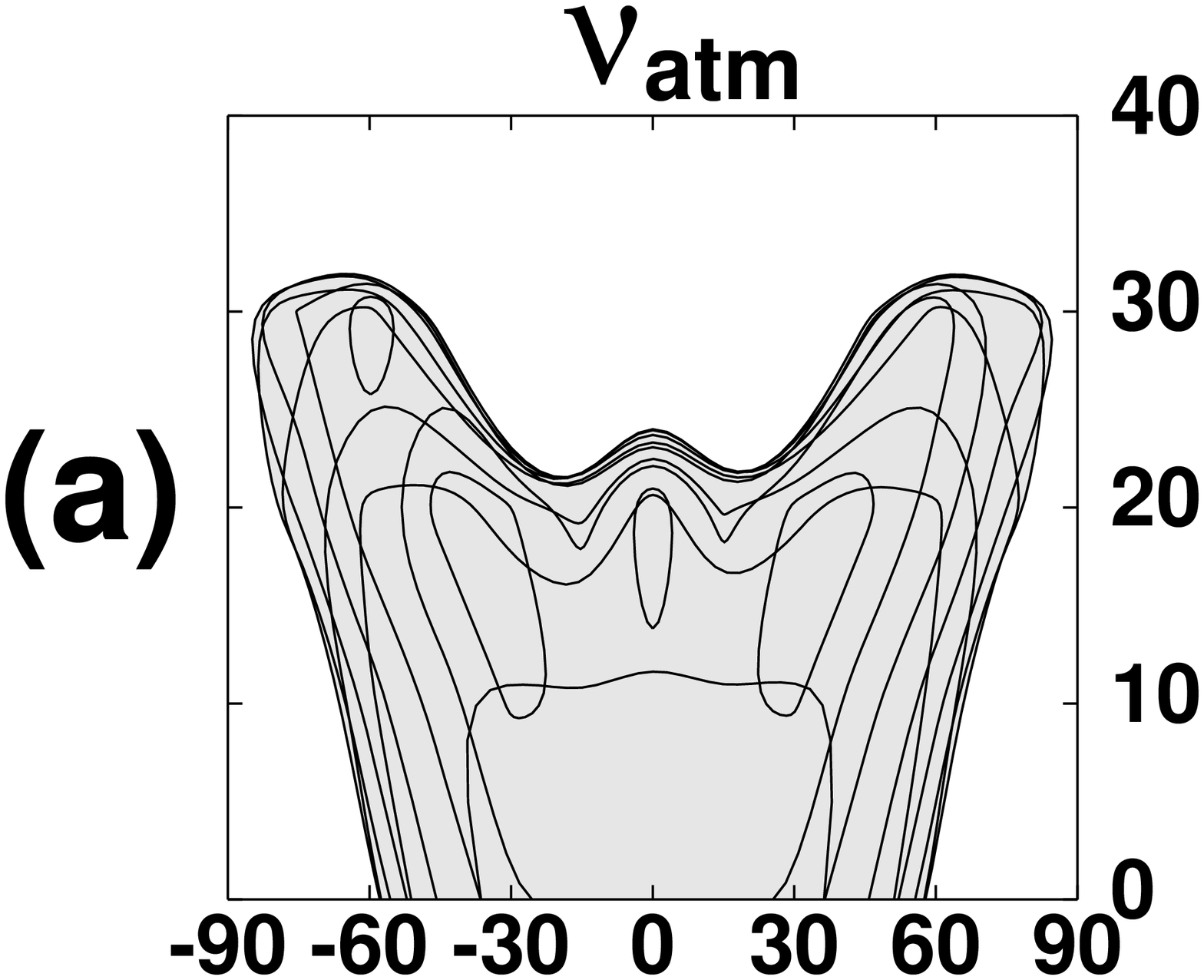}\hss}
\vglue -2.3cm
\hglue -2.5cm
\hbox to\hsize{\hss\epsfxsize=6.2cm\epsfbox{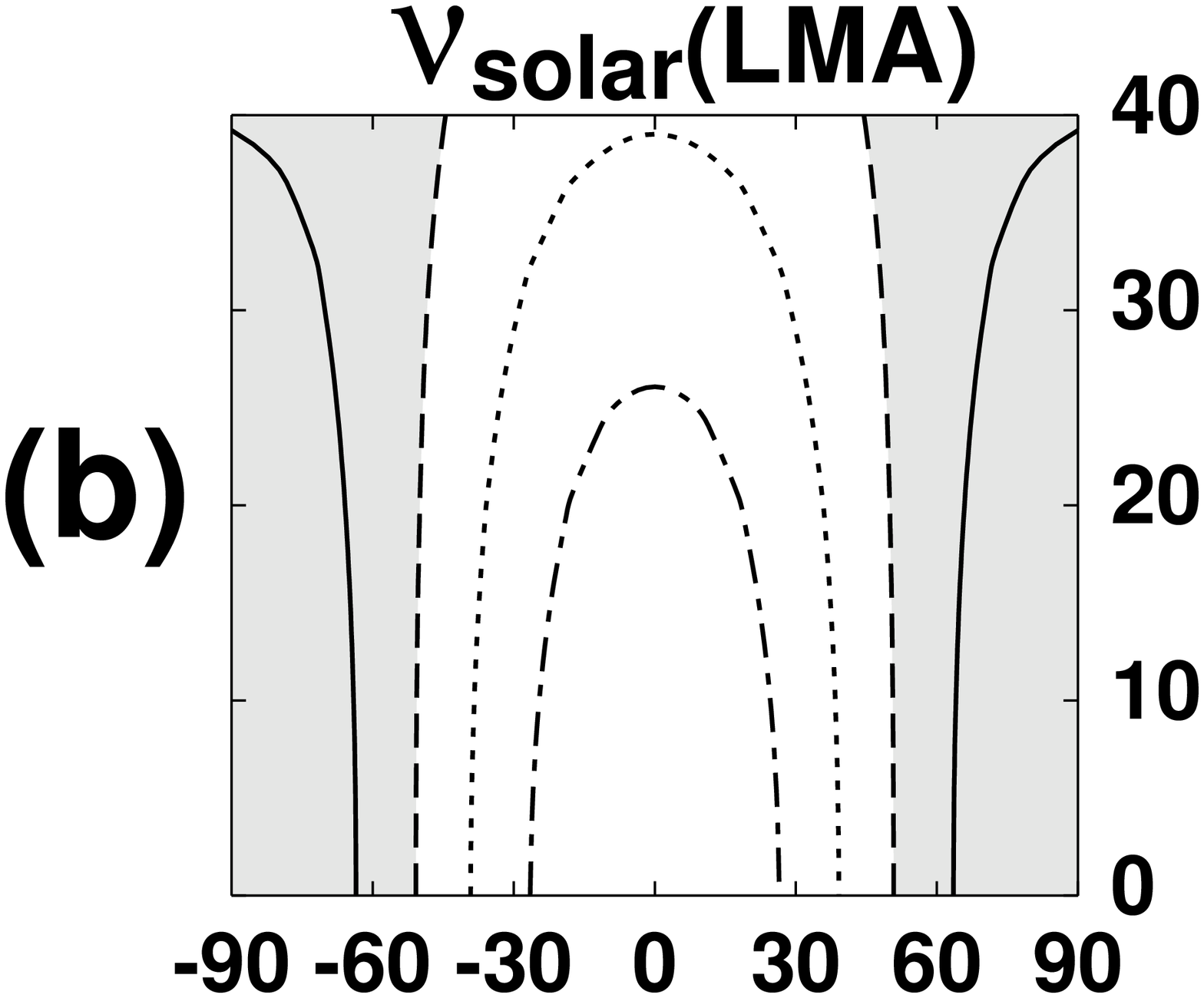}\hss}
\vglue -4.35cm
\hglue 0cm
\hbox to\hsize{\hss\epsfxsize=5.2cm\epsfbox{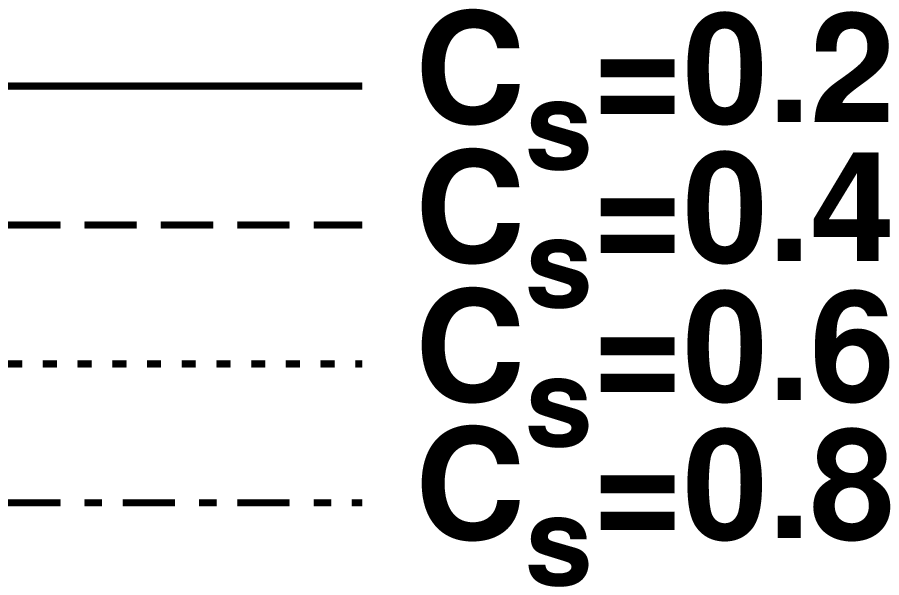}\hss}
\vglue -3.0cm
\hglue -2.5cm
\hbox to\hsize{\hss\epsfxsize=6.2cm\epsfbox{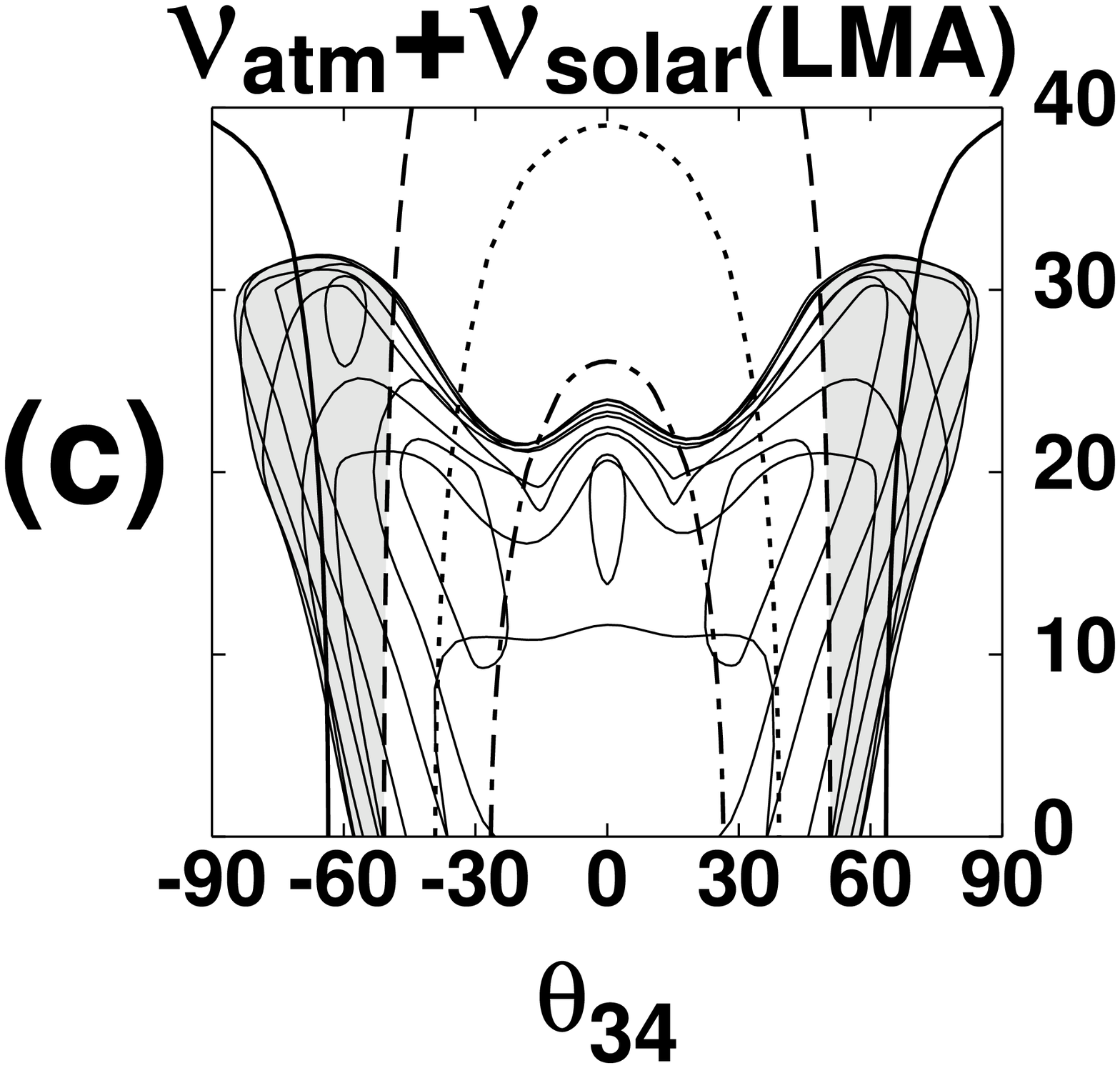}\hss}
\vglue -0.5cm
\caption{\label{fig:fig1}Allowed region for $\delta_1=\pi/2$,
$\theta_{24}=\pi/4$}
\end{figure}

Four neutrino mixing schemes have caught much interest, since they are
the simplest scenario which accounts for
the solar and atmospheric neutrino problems and the LSND data\cite{lsnd}
in the framework of neutrino oscillations.
To reconcile the data at 90\%CL of LSND,
Bugey and CDHSW one has to have two degenerate massive
states\cite{oy,bgg} ($m_{1}^2 \simeq m_{2}^2 \ll m_{3}^2 \simeq m_{4}^2$).
For simplicity I assume
$\Delta m_{21}^2=\Delta m_\odot^2$,
$\Delta m_{43}^2=\Delta m_{\mbox{\rm\scriptsize atm}}^2$
and I adopt the notation in \cite{oy} for the $4\times 4$ MNS matrix:
\begin{eqnarray}
\left( \begin{array}{c} \nu_e  \\ \nu_{\mu} \\ 
\nu_{\tau}\\\nu_s \end{array} \right)
=\left(
\begin{array}{cccc}
U_{e1} & U_{e2} &  U_{e3} &  U_{e4}\\
U_{\mu 1} & U_{\mu 2} & U_{\mu 3} & U_{\mu 4}\\
U_{\tau 1} & U_{\tau 2} & U_{\tau 3} & U_{\tau 4}\\
U_{s1} & U_{s2} &  U_{s3} &  U_{s4}
\end{array}\right)
\left( \begin{array}{c} \nu_1  \\ \nu_2 \\ 
\nu_3\\\nu_4 \end{array} \right).
\nonumber
\end{eqnarray}
For the range of the $\Delta m^{2}$ suggested by the LSND data at 90\%CL,
which is given by 0.2 eV$^2~\lsim
\Delta m^{2}_{\mbox{\rm{\scriptsize LSND}}} \lsim$ 2
eV$^2$ when combined with the data of Bugey E776 and KARMEN2, the
constraint by the Bugey data gives
$|U_{e3}|^2+|U_{e4}|^2\ll 1$.
Also $|U_{s3}|^2+|U_{s4}|^2$ has to be very small\cite{oy,bggs}
if one demands
that the number $N_\nu$ of effective neutrinos in Big Bang
Nucleosynthesis (BBN) be less than four.
In this case the solar neutrino deficit is explained by
$\nu_e\leftrightarrow\nu_s$ oscillations with the Small Mixing Angle
(SMA) MSW solution and the atmospheric neutrino anomaly is accounted
for by $\nu_\mu\leftrightarrow\nu_\tau$.  However, some people have
given conservative estimate for $N_\nu$ and if their estimate is
correct then $|U_{s3}|^2+|U_{s4}|^2\ll 1$ may no longer hold.
Recently the Superkamiokande group has reported
their result\cite{suzuki} on the
solar neutrino experiment which indicates that the
SMA and the Vacuum Oscillation solutions are disfavored.
Meanwhile
Giunti, Gonzalez-Garcia and Pe\~na-Garay\cite{ggp} have analyzed the
solar neutrino data in the four neutrino scheme without BBN
constraints.  Their updated results show that the SMA solution exists for
$0\le c_s \lsim 0.8$ ($c_s\equiv|U_{s1}|^2+|U_{s2}|^2$), while the
Large Mixing Angle (LMA) and LOW solutions survive only for $0\le c_s
\lsim 0.4$ and $0\le c_s \lsim 0.2$, respectively.
In this talk I will present some of the updated results of
my work\cite{y} on the four neutrino
oscillation analysis of the Superkamiokande atmospheric neutrino data
and I will briefly give some implications
to long baseline experiments, assuming that the solar neutrino deficit
is solved by the LMA solution.  For details of the analysis and
the references see \cite{y}.
In the analysis of atmospheric neutrinos, the effect of
$\Delta m_\odot^2$ is negligible, so I assume $\Delta m_{21}^2= 0$
and $U_{e3}=U_{e4}=0$ for simplicity.
To avoid contradiction with the CDHSW data, I will take $\Delta
m^{2}_{32}$=0.3eV$^2$ as a reference value.  The best fit to the
atmospheric neutrino data\cite{sobel} for 1144 days is obtained for $\Delta
m_{43}^2=2.0\times 10^{-3}{\mbox{\rm eV}}^2$,
$(\theta_{24},\theta_{34},\theta_{23})= (45^\circ,-30^\circ,20^\circ)$,
$\delta_1=45^\circ$ and the allowed region at 90\%CL is obtained
(The allowed region for 1144 day data is almost the same as for 990 day
data).
As a sample let me show the result for the case of
$\theta_{24}= 45^\circ,~\delta_1=90^\circ$.
The shadowed area in Fig.\ref{fig:fig1} (a) is the 90\%CL allowed region
projected on the
$(\theta_{34},~ \theta_{23})$ plane for various values of $\Delta
m_{43}^2$.
If I demand that the solar neutrino deficit be solved by the LMA
solution, then $c_s\lsim 0.4$ which is
depicted in Fig.\ref{fig:fig1} (b).  Combining the results
on the atmospheric neutrinos and the solar neutrinos,
the allowed region becomes the shadowed area in Fig.\ref{fig:fig1} (c).
It turns out that if I require $c_s\lsim 0.4$
then the solution prefers relatively large $\theta_{23}$
for any value of $\delta_1$.
\begin{figure}[h]
\vglue -1.0cm
\hglue 0.5cm
\hbox to\hsize{\hss\epsfxsize=6.2cm\epsfbox{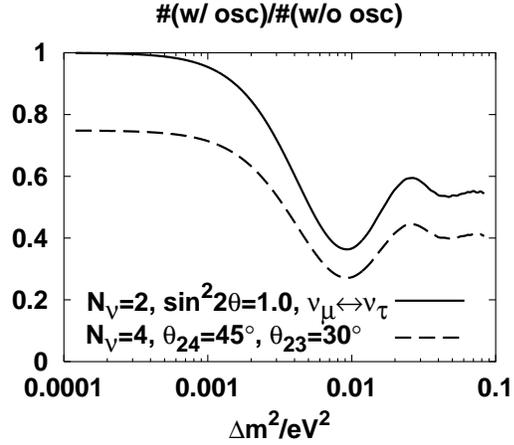}\hss}
\vglue 0.5cm
\caption{\label{fig:fig2}The ratio $r$ at K2K}
\end{figure}

\section{Application to long baseline experiments}
First implication of the present scheme is
the deficit which is expected to be seen at the K2K experiments.
Ignoring the matter effect,
I have for the neutrino energy $E\sim$ 1GeV and the path length $L$=250km
\begin{eqnarray}
P(\nu_\mu\rightarrow\nu_\mu)=1-2|U_{\mu 2}|^2(1-|U_{\mu 2}|^2)
\nonumber\\
-4|U_{\mu 3}|^2|U_{\mu 4}|^2
\sin^2\left({\Delta m_{\mbox{\rm\scriptsize atm}}^2L \over 4E}\right),
\label{eqn:k2kp}
\end{eqnarray}
where $|U_{\mu 2}|=|c_{24}s_{23}|$ is not necessarily small.
The behavior of the ratio 
$r\equiv$\#(CC + NC events with oscillations)/
\#(CC + NC events without oscillations) as a function of
$\Delta m_{\mbox{\rm\scriptsize atm}}^2$ is given in Fig.\ref{fig:fig2}.
Because of the contribution of the first term on the RHS of (\ref{eqn:k2kp})
the ratio is lower for the present four neutrino
scheme than for the standard two flavor case
with the same $\Delta m_{\mbox{\rm\scriptsize atm}}^2$.

\begin{figure}[h]
\vglue -1.0cm
\hglue 0.2cm
\hbox to\hsize{\hss\epsfxsize=6.2cm\epsfbox{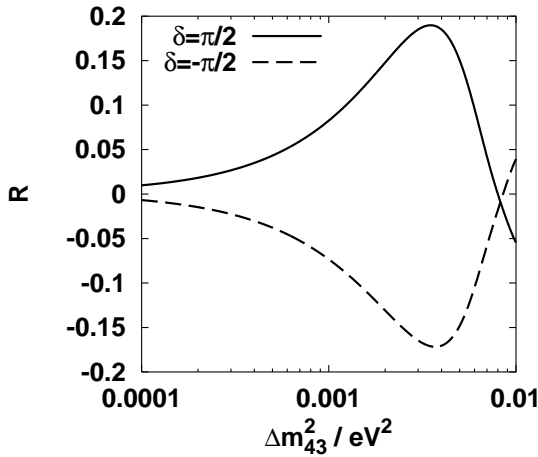}\hss}
\vglue 0.5cm
\caption{\label{fig:fig3}The ratio R at JHF}
\end{figure}

Second implication is the possible measurement of CP violation at the
JHF experiment\cite{jhf} which is expected to start from 2006.
Since there is no strong constraint on the mixing angles
$\theta_{24}$, $\theta_{34}$, $\theta_{23}$, CP violation in the channel
$\nu_\mu\rightarrow\nu_s$
could be large.
If we measure the neutral current $\pi^0$
production for $\nu_\mu$ and $\bar{\nu}_\mu$ beams and compare
them\cite{nakaya}, then the absolute value of the ratio
\begin{eqnarray}
R\equiv
{\left.{N_\nu(\delta_1) \over N_{\bar\nu}(\delta_1)}
\right|_{\mbox{\rm\scriptsize dat}}
-\left.{N_\nu(\delta_1=0) \over N_{\bar\nu}(\delta_1=0)}
\right|_{\mbox{\rm\scriptsize MC}}
\over
{\left.{N_\nu(\delta_1) \over N_{\bar\nu}(\delta_1)}
\right|_{\mbox{\rm\scriptsize dat}}
+\left.{N_\nu(\delta_1=0) \over
N_{\bar\nu}(\delta_1=0)}\right|_{\mbox{\rm\scriptsize MC}}}}
\nonumber
\end{eqnarray}
could be significantly larger than the
statistical fluctuation ($\sim$ 0.05 for $10^{21}$ POT)
(The yields is given in Table~\ref{tab:tab}
for an optimistic set of the oscillation parameters
($\delta_1=90^\circ$, $\theta_{24}=40^\circ$,
$\theta_{23}=30^\circ$, $\theta_{34}=60^\circ$,
$\Delta m_{\mbox{\rm\scriptsize atm}}^2=1.6\times 10^{-3}$eV$^2$).
The ratio $R$ is depicted in Fig.\ref{fig:fig3}
where the same set of the oscillation parameters as in Table~\ref{tab:tab}
and the Wide Band Beam\cite{jhf} is assumed in calculations).

\begin{table}
\caption{Yields of NC $\pi^0$ at JHF with $10^{21}$ POT}\label{tab:tab}
\begin{tabular}{|c|c|c|c|c|} 
 
\hline 
 
\raisebox{0pt}[12pt][6pt]{} & 
 
\raisebox{0pt}[12pt][6pt]{no osc} & 
 
\raisebox{0pt}[12pt][6pt]{$\delta_1={\pi \over 2}$} &
 
\raisebox{0pt}[12pt][6pt]{$\delta_1=0$} &
 
\raisebox{0pt}[12pt][6pt]{$\delta_1=-{\pi \over 2}$} \\
 
\hline
 
\raisebox{0pt}[12pt][6pt]{$N_\nu$} & 
 
\raisebox{0pt}[12pt][6pt]{393} & 
 
\raisebox{0pt}[12pt][6pt]{357} & 
 
\raisebox{0pt}[12pt][6pt]{311} & 
 
\raisebox{0pt}[12pt][6pt]{282} \\

\hline

\raisebox{0pt}[12pt][6pt]{$N_{\bar\nu}$} & 
 
\raisebox{0pt}[12pt][6pt]{199} & 
 
\raisebox{0pt}[12pt][6pt]{146} & 
 
\raisebox{0pt}[12pt][6pt]{158} & 
 
\raisebox{0pt}[12pt][6pt]{184} \\\hline
\end{tabular}
\end{table}

\section{Conclusions}
I have presented my result on the
four neutrino oscillations of
the Superkamiokande atmospheric neutrino data
without assuming the BBN constraints.
By combining the analysis on the solar
neutrino data by Giunti et al. and assuming the LMA solar solution,
I found that there is relatively large contribution of
$\Delta m_{\mbox{\rm\scriptsize LSND}}^2$ in the atmospheric neutrino
oscillations.  It suggests that the number of
events at K2K is less than the one for the standard two flavor
scenario for the same $\Delta m_{\mbox{\rm\scriptsize atm}}^2$
and that CP violation could be measured at JHF
after running for several years with $10^{21}$POT/yr.

\section*{Acknowledgments}
I would like to thank T. Nakaya for pointing out
the NC $\pi^0$ channel to me to measure CP violation.
This
research was supported in part by a Grant-in-Aid for Scientific
Research of the Ministry of Education, Science and Culture,
\#12047222, \#10640280.

\end{document}